\documentclass[traditabstract]{aa} 
\usepackage{natbib}
\bibpunct{(}{)}{;}{a}{}{,}
\usepackage{graphicx}
\usepackage{savesym}
\usepackage{amsmath}
\savesymbol{iint}
\usepackage{txfonts}
\restoresymbol{TXF}{iint}
\usepackage[dvipdfm,hyperfootnotes=false]{hyperref}
\usepackage{amsmath,amssymb,amsfonts,textcomp}
\usepackage{color}
\usepackage{subfig}
\usepackage{float}
\usepackage{verbatim}
\newcommand{\msol}{$M_{\odot}$}
\newcommand{\mbh}{M_{\bullet}}
\newcommand{\bildan}[4]{
  \begin{figure}[#1]
    \resizebox{\hsize}{!}{\includegraphics{#2}}
    \caption{#3}
    \label{#4}
  }
\newcommand{\bildaus}{\end{figure}}
\newcommand{\bildangross}[4]{
  \begin{figure*}[#1]
    \centering
    \includegraphics[width=17cm]{#2}
    \caption{#3}
    \label{#4}
  }
\newcommand{\bildausgross}{\end{figure*}}
\newcommand{\tableOn}[4]{
  \begin{table}[#1]
    \caption{#2}
    \label{#3}
    \centering
    \begin{tabular}{#4}
      \hline \hline
    }
\newcommand{\tableOff}{\hline
			\end{tabular}
			 \end{table}
                       }
\renewcommand{\comment}[1]{}

\newcommand{\tsup}[1]{\textsuperscript{#1}}
\begin{document}
   \title{No evidence for a central IMBH in M15}

   \author{F. Kirsten 
          \inst{1}\fnmsep\inst{2}\fnmsep\thanks{franz@astro.uni-bonn.de}
          \and
          W.H.T. Vlemmings\inst{3}
	}
   \institute{Argelander Institut f\"ur Astronomie (AIfA), University of Bonn,
              Auf dem H\"ugel 71, D-53121 Bonn, Germany
         \and
              Max Planck Institut f\"ur Radioastronomie (MPIfR), 
              Auf dem H\"ugel 69, D-53121 Bonn, Germany             
	\and	
        Department of Earth and Space Sciences, Chalmers University of
        Technology, Onsala Space Observatory, SE-439 92 Onsala, Sweden
             }

   \date{}

 
  \abstract
   {Intermediate mass black holes (IMBHs) with expected masses $\mbh\approx
     10^{4}$ \mbox{\msol} are thought to bridge the gap between stellar mass black
     holes ($\mbh\approx 3 - 100$ \msol) and supermassive black holes found at
     the centre of galaxies ($\mbh > 10^{6}$ \msol). Until today, no IMBH
     has been confirmed observationally. The most promising objects to host an
     IMBH as their central mass are globular clusters.  Here, we present high
     sensitivity 
     multi-epoch 1.6 GHz very long baseline interferometry observations of the
     globular cluster M15 that has been suggested to host an IMBH. Assuming
     the IMBH to be accreting matter from its surrounding we expect to detect
     it as a point source moving with the global motion of the
     cluster. However, we do not detect any such object within a radius of
     6000 AU of the cluster centre in any of the five observations
     spread over more than one year. This rules out any variability of the putative IMBH on the time scale of one to two months.  To get the most stringent upper limit
       for the flux density of the putative IMBH we concatenate the data of all five
       epochs. In this data we measure a 3$\sigma$ upper
     flux limit of 10 $\mu$Jy for a central source. We employ the fundamental
     plane of black hole activity to estimate the mass of the central IMBH
     candidate. Based on previous X-ray observations of M15 our measurements
     indicate a 3$\sigma$ upper mass limit of $\approx500$ \msol.
}


\maketitle
%
%

\section{Introduction}
Super-massive black holes (SMBHs) with masses $\mbh \approx 10^{6-9}$
solar masses (\mbox{\msol}) are known to exist at the centre of galaxies
\citep{kormendy95}. At the same time, observations of high-mass X-ray binaries led to the conclusion that stellar-mass black holes with $\mbh \approx 100$ \mbox{\msol} must form as
well \citep{mcclintock06,ozel10}. The existence of black holes (BHs) bridging the gap between these two extremes in
mass, the so-called intermediate mass black holes (IMBHs, $\mbh \approx 10^{4}$ \mbox{\msol}), is still under
debate. 

One possible candidate for IMBHs are ultraluminous X-ray sources (ULX,
\citeauthor{colbert99} \citeyear{colbert99}) appearing to accrete matter at
super-eddington rates. The interpretation of ULX--properties as being
characteristic for IMBHs,
however, is still in discussion (e.g. \citeauthor{berghea08} 2008,
\citeauthor{zampieri09} 2009). Nevertheless, the recent discovery of the hyper-luminous X-ray
source ESO 243-49 HLX-1 by \citet{farrell09} adds evidence that ULXs might
host IMBHs.

Based on the black hole mass--stellar velocity dispersion relation ($\mbh -\sigma$) established for galaxies \citep{ferrarese00,gebhardt00}, an object
like an IMBH can be expected to reside at the core of globular
clusters (GCs). Even though recent work by \citet{vesperini10} shows that an IMBH can be
formed in a GC environment, no conclusive observational evidence for the existence of
IMBHs has been found to date.

Apart from, e.g, the globular clusters $\omega$ Cen \citep{vandermarel10}, G1 in M31
\citep{gebhardt02, ulvestad07}, and 47 Tuc \citep{lu11}, M15 has been one of the most promising GC--candidates to host an
IMBH for a long time. Stellar surface density profiles of M15 reveal a
steady increase towards the center of the cluster indicating a state of
advanced core-collapse \citep{djorgovski86}. To explain the central brightness peak \citet{newell76} suggested an 800 \msol black hole to
reside at the core of the cluster. The increase of the stellar
velocity dispersion towards the centre of M15 as measured by, e.g, \citet{gerssen03}, supports the notion of a high central mass
concentration in the cluster. In fact, dynamical models based
on line-of-sight velocities and proper motions infer a mass of 3400 \mbox{\msol} within the central 1 arcsecond ($= 0.05$ \mbox{pc} at the distance of
$10.3 \pm 0.4$ \mbox{kpc}, \citeauthor{vandenbosch06} 2006). The nature of this mass concentration is
unknown. \citet{gerssen03} invoke the existence of an
IMBH with a mass of $\mbh=1700^{+2700}_{-1700}$
\mbox{\msol} to explain their observations. Similarly to
\citet{illingworth77}, \citet{baumgardt03} and \citet{mcnamara03} challenge this interpretation, based on N-body simulations
excluding an IMBH. Instead, these authors claim that the
observational data can be explained equally well by assuming a collection of neutron stars to
exist at the core of M15. The latest simulations of that kind
require a total of 1600 neutron stars to fit the velocity-dispersion profile
mentioned above \citep{murphy11}. 

Complementary to the indirect (non-)evidence
using kinematic studies based on optical observations, X-ray and radio
observations aim at directly detecting such an object. The fundamental
  plane of black hole activity (FP)
as determined for active galactic nuclei (AGN) \citep{merloni03, falcke04}, relates
black-hole mass, X-ray and radio luminosity. Now, assuming the same
physical processes powering AGN--emission to also be characteristic for IMBHs,
\citet{maccarone04} predict a black hole mass of 400 \mbox{\msol} for the IMBH
in M15. Their result is based on an estimate of the cluster's total mass,
$M_{\text{GC}}$, from its absolute V-magnitude, $M_{\text{V}}$, and on a model
by \citealt{miller02} stating that $M_{\bullet}\approx
10^{-3}M_{\text{GC}}$. \citet{bash08}, on the other hand, perform a survey of M15 at 8.6 GHz using
the Very Large Array that reaches a noise level of 8.5
\mbox{${\mu}$Jy/Beam}. Using the FP they predict a flux density of
$10^{3}-10^{5}$ \mbox{$\mu$Jy} for a putative IMBH with a spectral index
$\alpha=-0.7$. They detect no central source at
a $3\sigma$ upper flux limit of 25 \mbox{$\mu$Jy}. Similary, \citet{cseh10}
tried to detect an IMBH at the center of the globular cluster NGC 6388. The $3\sigma$
noise level (81 \mbox{$\mu$Jy}) of their observations with the Australia
Telescope Compact Array allowed the authors to constrain the mass of the
possible IMBH to be lower than $\approx1500$ \msol.

In this paper we discuss multi-epoch observations of M15 almost three times as
sensitive as those of \citet{bash08}. The high angular resolution of our data
allows us to disentangle any possible background sources from objects
belonging to the cluster. Furthermore, the long time line spanning 15 months
allows us to, in principle, detect the proper motion of a possible central radio
source moving with the global motion of the cluster expected to be on the
order of $-1.0\pm0.4$ and $-3.6\pm0.8$ \mbox{mas/yr} in right ascension (RA)
and declination (Dec), respectively \citep{jacoby06}.
\section{Observations}
We observed M15 five times in a global VLBI campaign that was spread over a
time period of more than one year. The observations included in this analysis were
conducted on 11 November 2009, 7 March 2010, 5 June 2010, 2 November
2010, and 27 February 2011. The array we employed consisted of eight European VLBI Network (EVN)
antennas (Jodrell Bank, Onsala, Westerbork, Effelsberg, Noto, Medicina,
Toru\'n, Arecibo) and the Greenbank Telescope (GBT). We observed at a central
frequency of 1.6 GHz and the data was recorded at 1024 Mbps. Accounting for the different receiver
systems at the individual telescopes our total bandwidth amounts to
230 MHz on average. The correlation was done at the EVN-MkIV correlator \citep{schilizzi01} at the  \textit{Joint
Institute for VLBI in Europe} (JIVE). 

The longest baselines in east-west (north-south) direction extending
over 7500 (2000) km allow for a resolution of $2.2\times6.3$ \mbox{mas}. The largest dishes
of the array (Arecibo, Effelsberg, and the GBT) ensure a maximum sensitivity
of approximately 4 \mbox{$\mu$Jy/Beam}.

The observing schedule lasted six hours in total, 3.6 hr of which were
spent on the target cluster M15. The quasar J2139+1423 (located
  $\approx3.17\degr$ to the north-east of the pointing center) served as phase
calibrator and the blazar 3C454.3 was used for bandpass
calibration. Arecibo-data is available for 75 (50) min in epochs 1 and 5
(epoch 3). Unfortunately, epochs 2 and 4 lack any Arecibo data which is why
the sensitivity and astrometric precision of these two datasets is lower by about a factor of two. 

Aiming to detect compact radio sources close to the core of M15 we map out the
entire central region within 2\mbox{\arcmin} in only one pointing. For this
project, however, only the very central region (the central 16\arcsec) correlated at RA = 21\tsup{h}29\tsup{s}58\fs3120, Dec =
12\degr10\arcmin02\farcs679 (J2000 equinox) is of interest. The entire dataset will be
described in a forthcoming paper.
\section{Data reduction}
After correlation, all data is reduced, calibrated and imaged using the NRAO
 \textit{Astronomical Image Processing System}
(AIPS\footnote{\url{http://www.aips.nrao.edu/}}). A priory calibration tables
including system temperature and gain curve corrections as well as a flag
table containing information about band edges and off-source times are
provided by the EVN
pipeline\footnote{\url{http://www.evlbi.org/pipeline/user\_expts.html}}. We
apply these to the dataset as given. Parallactic angle
corrections are determined with the AIPS task CLCOR and first ionospheric
corrections are computed running TECOR with the total electron content (TEC)
maps published by the \textit{Center for Orbit Determination in
  Europe}\footnote{\url{ftp://ftp.unibe.ch/aiub/CODE/}}. Even
though these maps are quite crude in angular resolution (about
$5\degr\times2.5\degr$) they have shown to be of use reducing the
scatter in phase delay by a factor of 2-5 \citep{walker99}. Next, we identify and flag radio frequency interference (RFI) for all antennas and sub bands.

The bandpass calibration is done running BPASS on the data for 3C454.3 and
yields phase and amplitude gain factors for all 8$\times$128 (512, epoch 1)
channels for all antennas. We align phases in between IFs by performing a manual fringe correction
running FRING on 3C454.3 on a sub-interval of about 30 s of
observation.

At this point, we combine all correction tables obtained so far and apply it
to the data of the phase calibrator. We fringe fit this dataset including data
over the entire time range. We solve for phase delays and phase rates
simultaneously using solution intervals of 1.5 min.

The fringe solutions in conjunction with all calibration solutions found
earlier are then applied to the M15 data. In order to eliminate any residual
phase delays and amplitude errors caused by the atmosphere and the ionosphere
we take advantage of the strong unclassified source S1 \citep{johnston91}
located about $94\arcsec$ to the west of the cluster centre and use it for in-beam calibration. To speed up the self-calibration process we average the
visibilities both in the time- and frequency domain to 2 s integration
time and 64 channels per IF. 

Finally, we image the self-calibrated data running IMAGR employing natural
weighting to ensure maximal sensitivity. In order to account for
possible inaccuracies in the assumed cluster centre we produce an image that
has an angular size of roughly $16\arcsec\times16\arcsec$ ($=0.8\times0.8$ \mbox{pc}).
\section{Analysis and results}
In all five epochs we first produce a noise map of the image by smoothing
it with a kernel that has a size of $1024\times1024$ pixels. Based on this rms
map, the AIPS source detection algorithm SAD then searches for objects down to a
signal-to-noise ratio of 3. The coordinates of the possible sources detected in
this fashion in all five observation epochs are then
cross-correlated. Cross-correlation is performed allowing
for a maximal positional shift of 15 mas in between epochs. This corresponds to roughly four times
the maximal beam width in right ascension. No match can be found relating all
five data sets. 

\begin{figure*}
  \centering
  \subfloat{\label{fig: imbh e1}
    \includegraphics[width=0.4\textwidth]{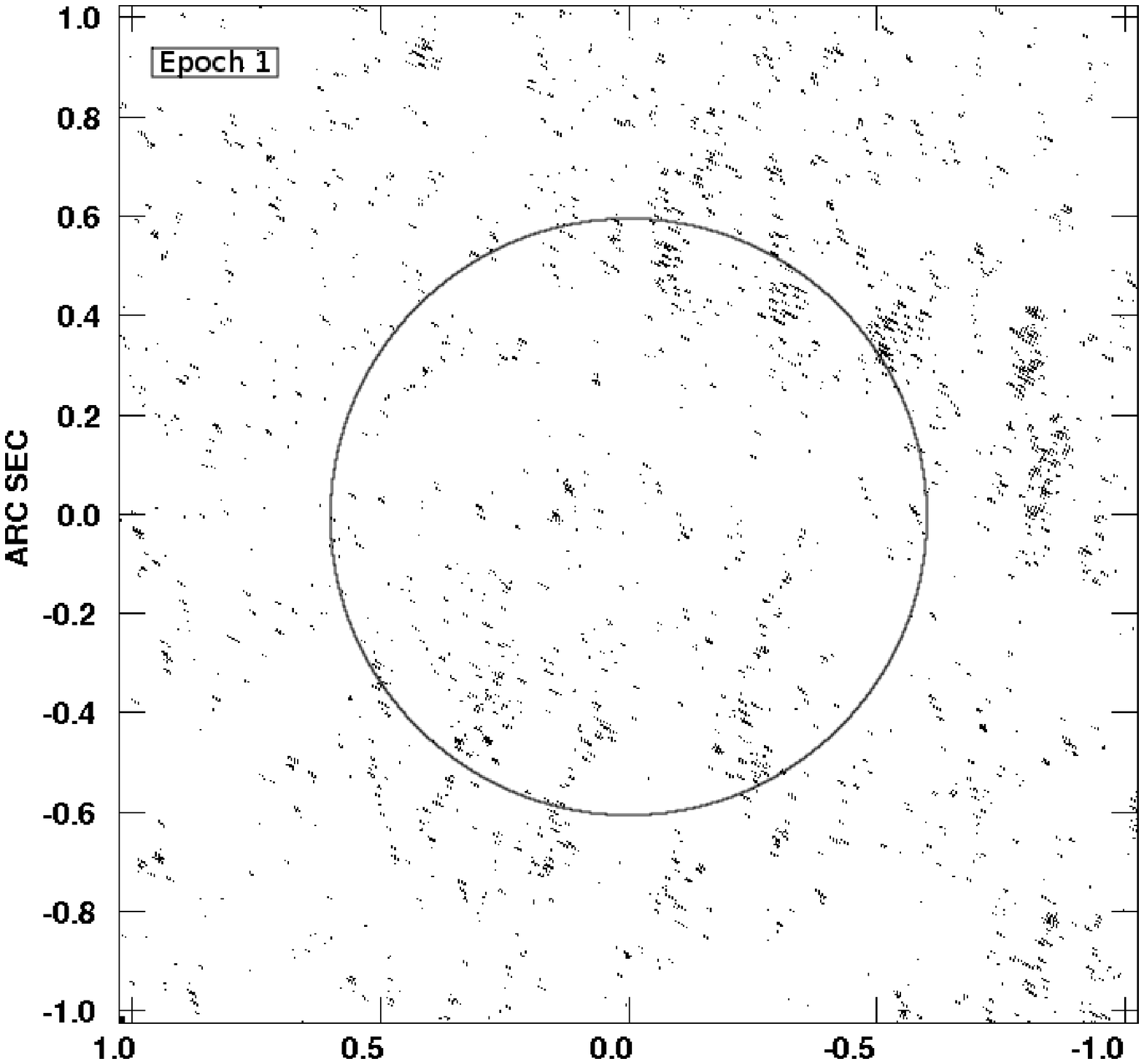}}  
  \subfloat{\label{fig: imbh e2}
    \includegraphics[width=0.4\textwidth]{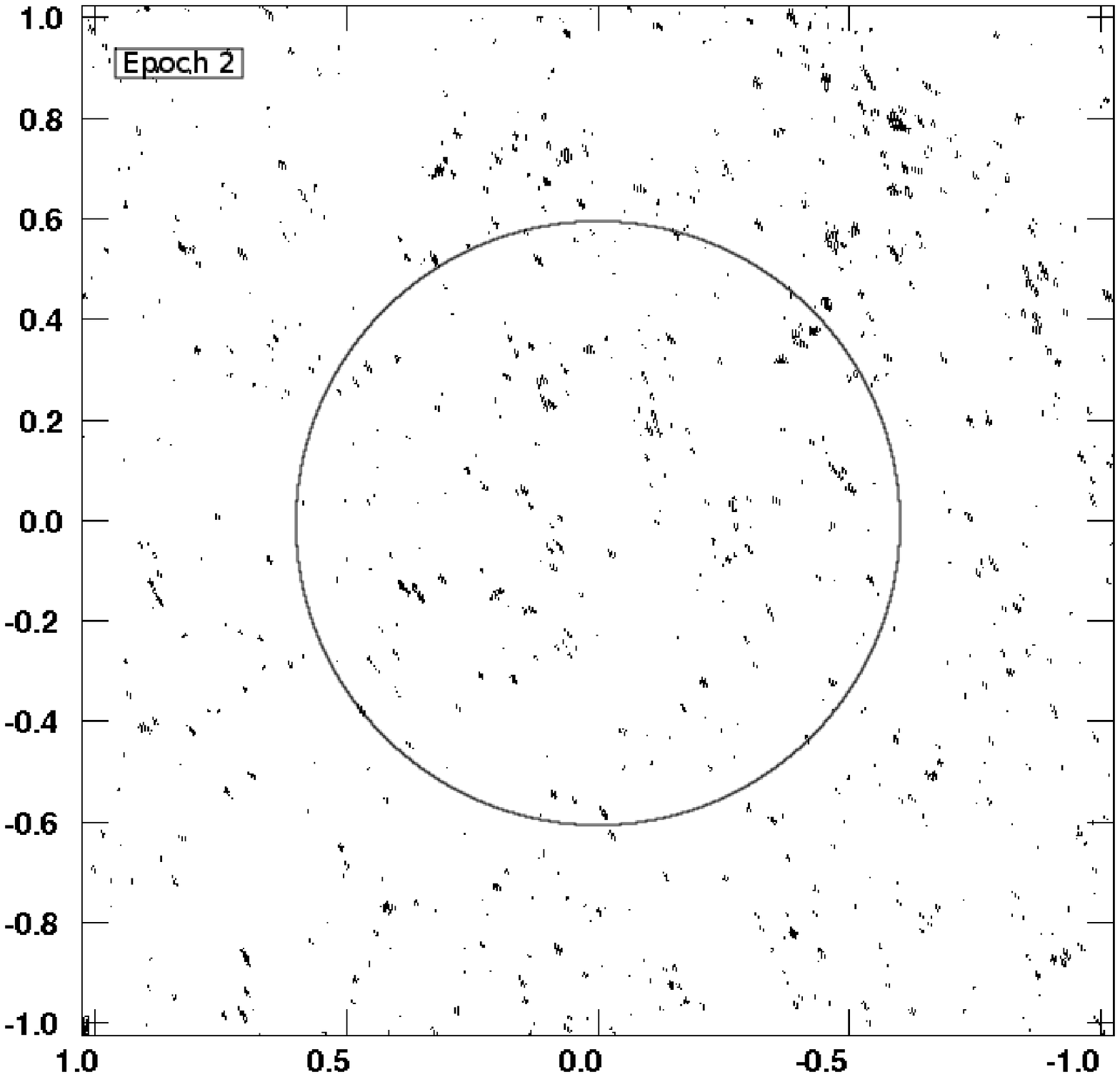}}\\
  \subfloat{\label{fig: imbh e3}
    \includegraphics[width=0.4\textwidth]{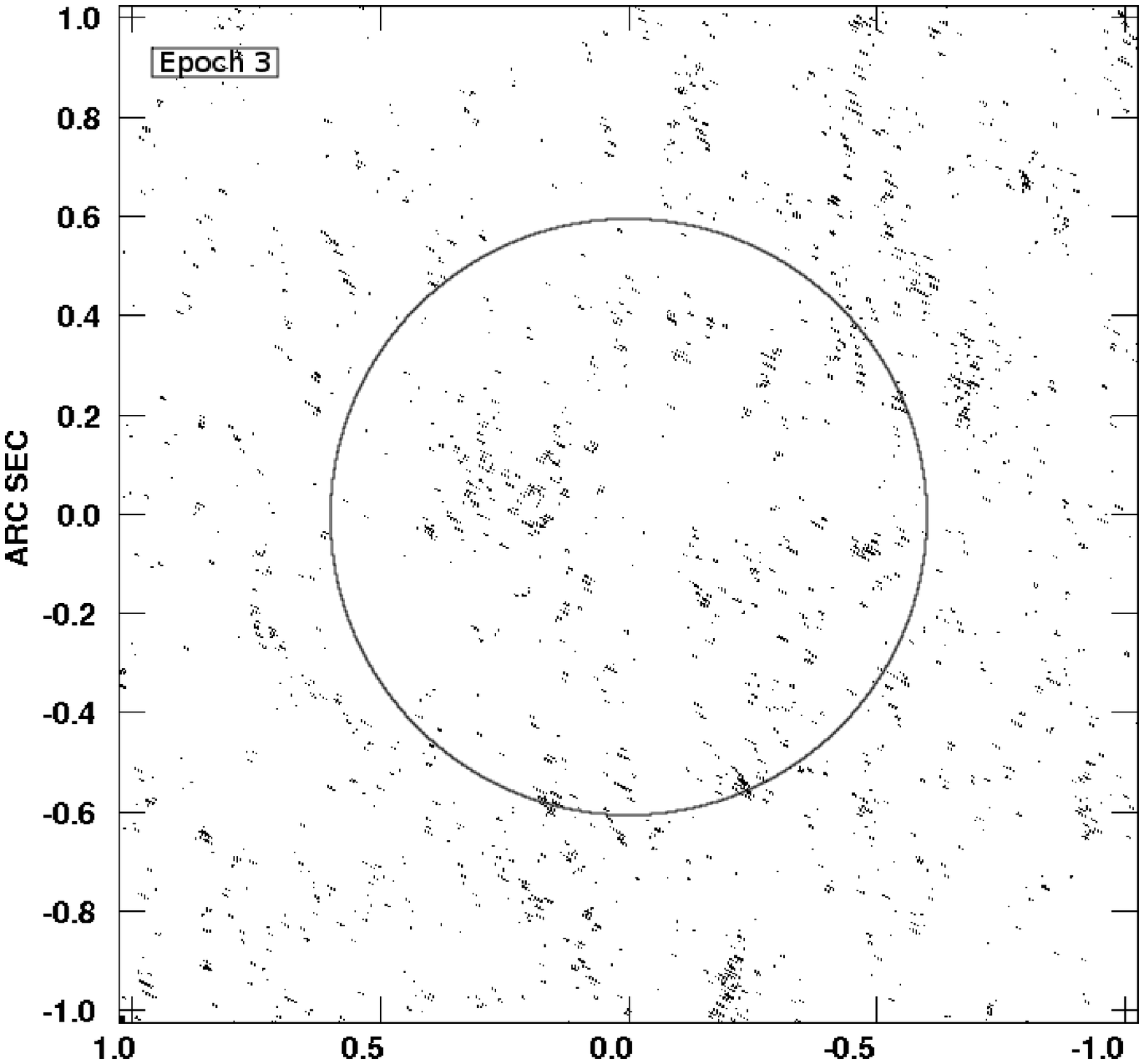}}
  \subfloat{\label{fig: imbh e4}
    \includegraphics[width=0.4\textwidth]{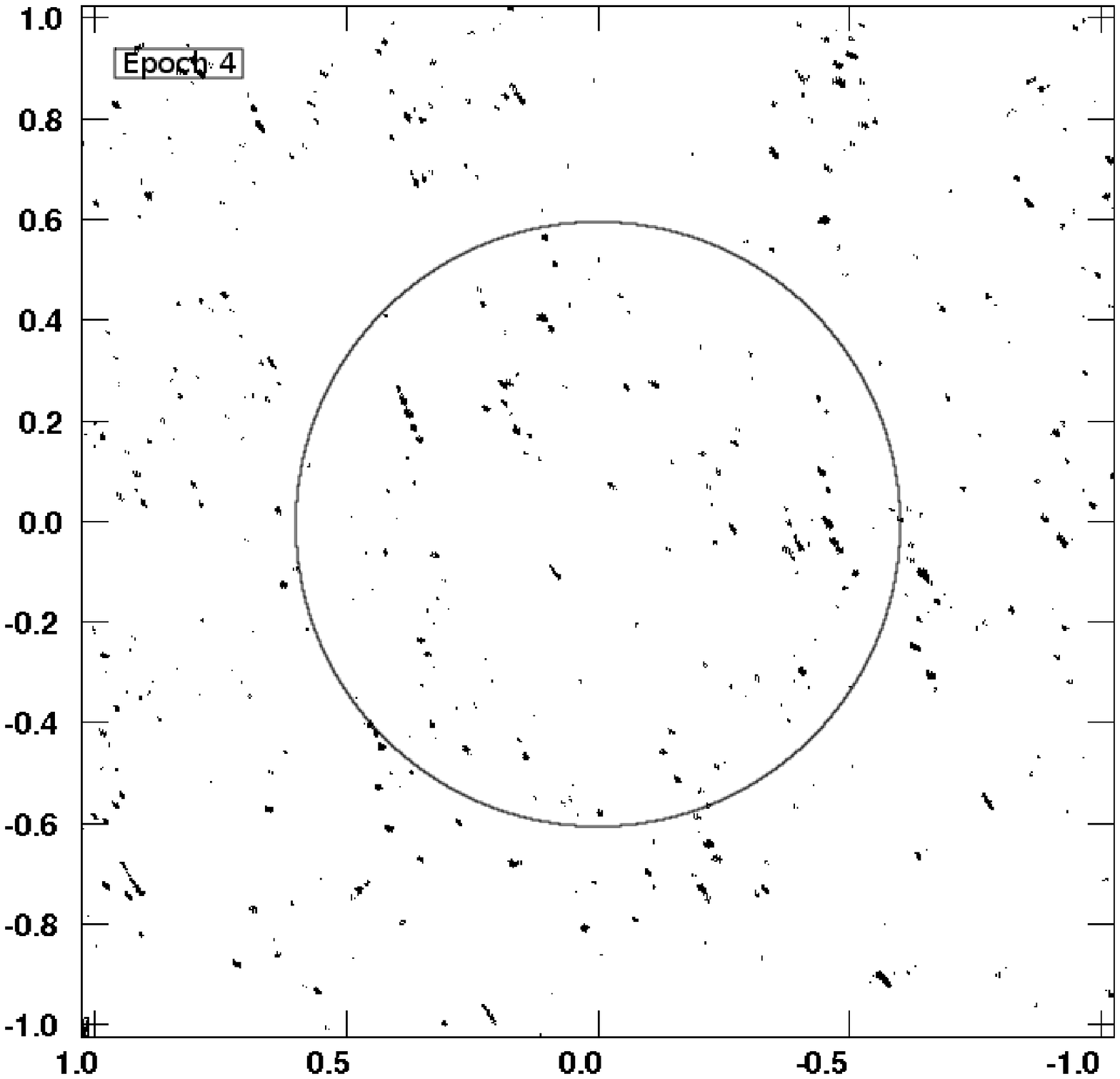}}\\
 \subfloat{\label{fig: imbh e5}
    \includegraphics[width=0.4\textwidth]{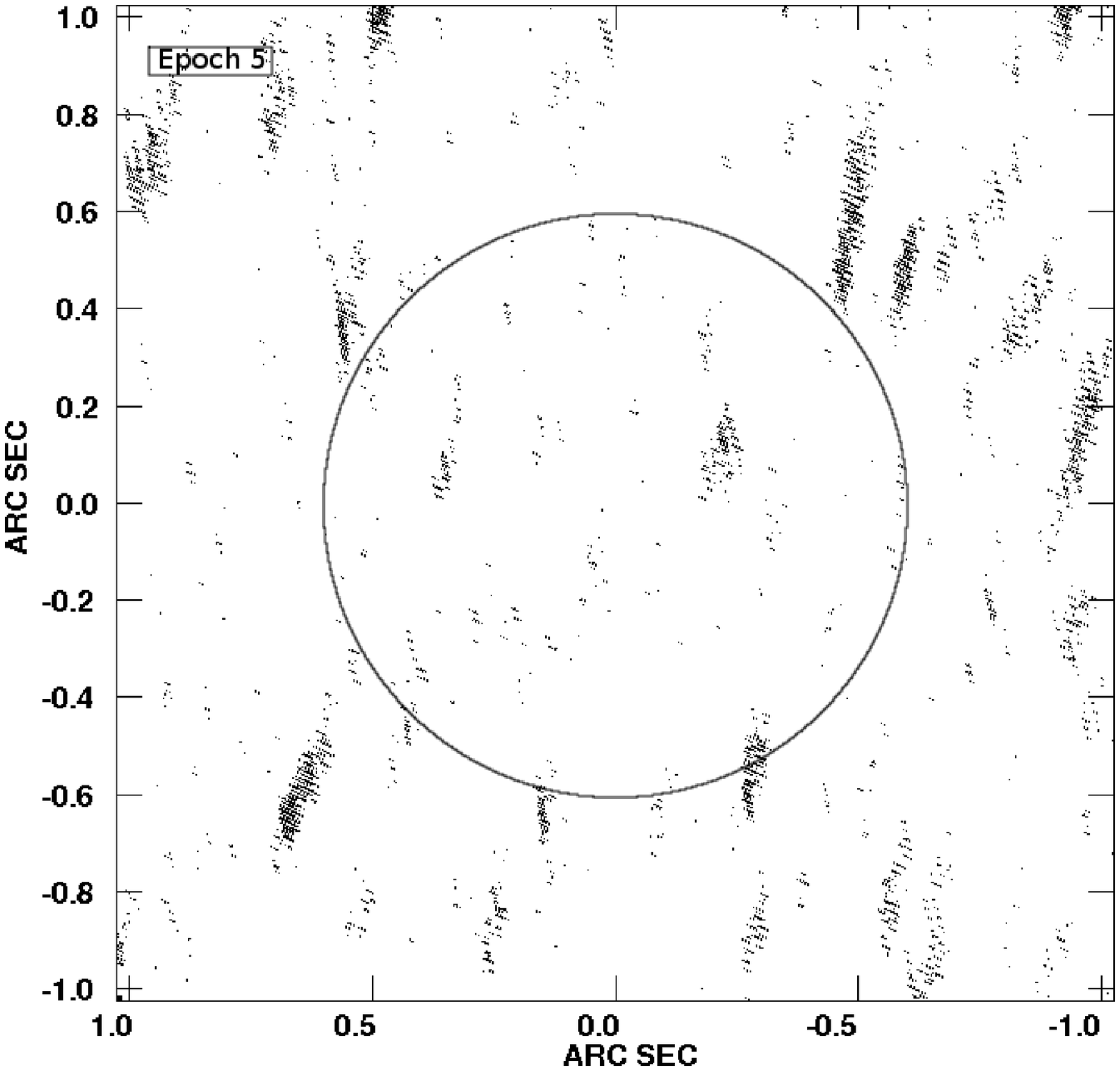}}
  \subfloat{\label{fig: imbh composite}
    \includegraphics[width=0.4\textwidth]{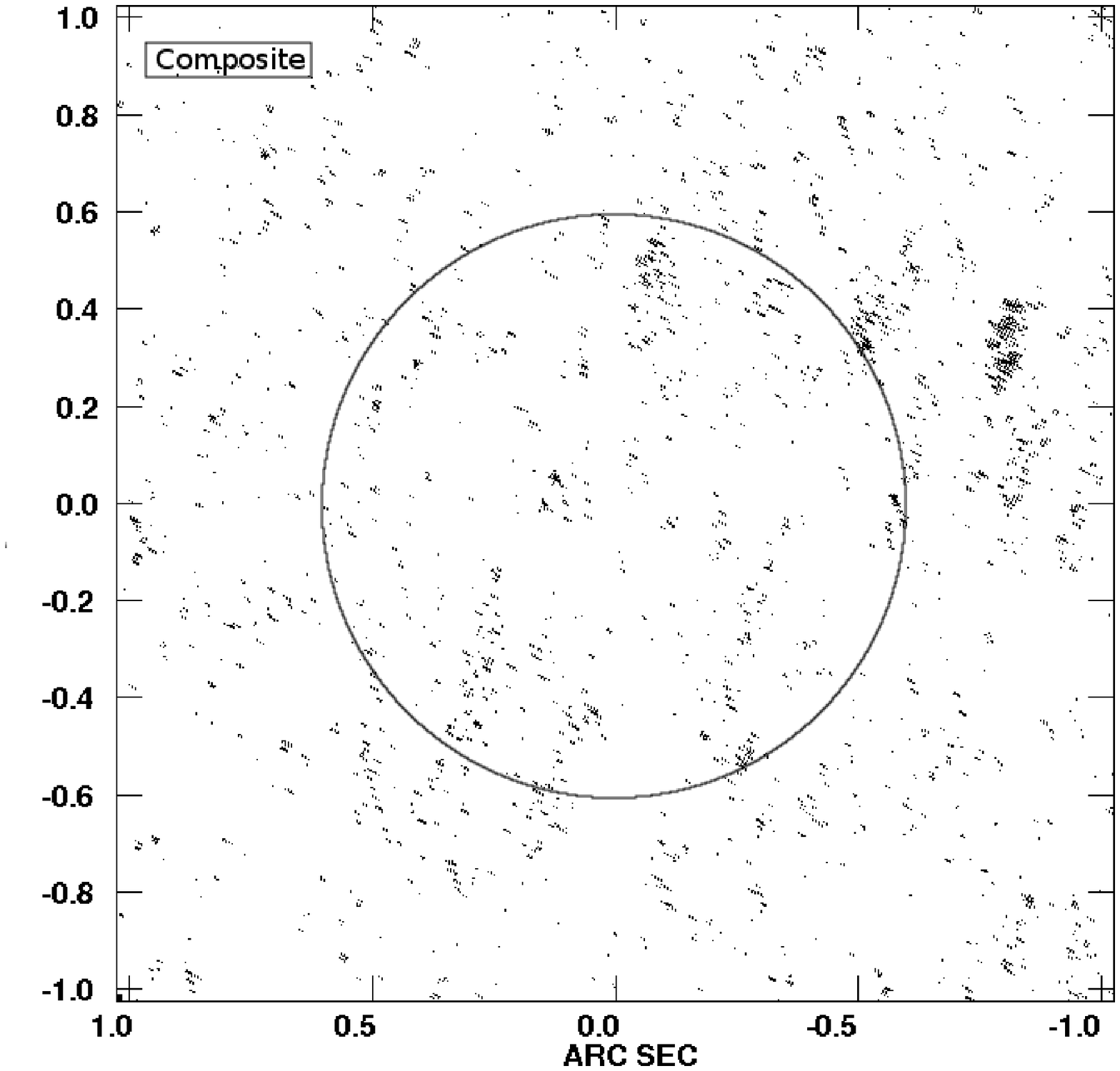}}
  \caption{Contour plots of all five individual observations and the composite
    image combining the data of all epochs. The tiles are centred on the assumed
    core of M15 at coordinates RA = 21\tsup{h}29\tsup{s}58\fs330, Dec =
12\degr10\arcmin01\farcs200. The circle indicates the 3$\sigma$ error of the
position of the core as determined by \citet{goldsbury10}. Adopting a distance
of 10.3 \mbox{kpc} to M15, each tile has a physical dimension of  about $(20
000$ \mbox{AU})$^{2}$. Contours are $(-5, -3, 3, 5)$ times the rms in each
individual epoch (4.7, 8.9, 4.3, 11.5, 5.8 \mbox{$\mu$Jy}). The noise level of the composite
image is 3.3 \mbox{$\mu$Jy}. }
\label{fig: imbh all}
\end{figure*}

Finally, we also inspect the images manually. Figure \ref{fig: imbh all}
displays contour plots of the dirty images of the central region of M15
for all five epochs. All tiles are centred on the cluster core as published
by \citet{goldsbury10} at coordinates RA = 21\tsup{h}29\tsup{s}58\fs330, Dec =
12\degr10\arcmin01\farcs200 (accurate to within 0.2\arcsec). We do not detect
a significant signal in any of the epochs. Accounting for the different
sensitivity limits varying between 4.3 \mbox{$\mu$Jy/Beam}  in epoch 3 and
11.5 \mbox{$\mu$Jy/Beam} in epoch 4, we can put upper 3$\sigma$ limits between 13
and 35 \mbox{$\mu$Jy/Beam} on the flux density of a possible central
object. Figure \ref{fig: imbh all} also displays
a dirty image of the data concatenated over all five epochs. The rms of this
deconvolved image is 3.3 \mbox{$\mu$Jy/Beam} which translates to a 3$\sigma$ upper flux limit of 10 \mbox{$\mu$Jy/Beam}. 
\section{Discussion}
\subsection{Radio flux limits and variability}
The five individual epochs have rather different sensitivity limits that correspond to a $3\sigma$ upper flux density limit of 14.1, 26.7, 12.9, 34.5, 17.4 $\mu$Jy/beam
  for epochs 1 to 5. We did not detect a central source in any of the observations that were conducted at regular intervals of about three months spanning a time range of 15
  months. Provided the IMBH-candiate is of transient nature that is in its 'on'--state for longer than one or two months, the probability to have missed it in all five observations is
  negligible. Therefore, we rule out any variability of the central object on these time scales. Accordingly, we can assume a steady
state emission model in which case the noise level of the concatenated data yields
the most stringent $3\sigma$ upper flux limit of $\approx10$ \mbox{$\mu$Jy}. This flux limit is a factor of 2.5 lower than that from \citet{bash08} and we will adopt it throughout the following analysis.

\subsection{Mass limits from X-ray observations}
Similarly to the first versions of the FP from \citet{merloni03} and \citet{falcke04}, the FP derived by \citet{kording06} uses a sample including both X-ray
binaries (XRBs) and AGN. Thus, both versions span several orders of magnitude in BH mass and should also be applicable for the intermediate mass range of IMBHs. The relation
found by \citet{kording06} for XRBs and low-luminosity radiatively inefficient AGN has the lowest intrinsic scatter $\sigma_{\text{int}}=0.12$ \mbox{dex}
($\approx$30 percent) and we will use it in the following. In terms of black hole mass $M_{\bullet}$, radio luminosity $L_R$, and X-ray luminosity $L_X$ their FP-relation reads
\[
\text{log}M_{\bullet}=1.55\;\text{log}L_R-0.98\;\text{log}L_X-9.95 \;\;.
\]
M15 is known to host two strong X-ray sources, AC211 \citep{giacconi74, clark75} and M15 X2 \citep{white01}, close to the core of the cluster. Both have been classified as low mass XRBs. The strong X-ray emission
of these two objects makes it difficult to detect the expectedly faint emission of a central IMBH. Nevertheless, \citet{ho03} put an upper limit of $L_X=5.6\times10^{32}$ \mbox{erg/s} on the X-ray
luminosity of the putative IMBH. \citet{hannikainen05} manage to detect a faint source ($L_X=3.3\times10^{32}$ \mbox{erg/s}) close to core of M15 which they attribute to a dwarf
nova. In their paper, \citet{hannikainen05} report a flux detection limit of $2\times10^{-15}$ \mbox{erg/s/cm} (0.5-2 keV) which translates to an X-ray luminosity of
$2.54\times10^{31}$ \mbox{erg/s} at the distance of M15. Converting our measured radio flux to a 5 GHz radio luminosity $L_R\leq 6.34\times10^{27}$ \mbox{erg/s} (assuming a flat
radio spectrum) and inserting it together with the upper limit for the X-ray luminosity from \citet{hannikainen05} the FP
yields a mass limit of $232\pm93$ \mbox{\msol}. This translates to a very conservative $3\sigma$ upper mass limit of 511 \mbox{\msol} for the putative IMBH. Our observations thus
decreased the upper mass limit by a factor of
four compared to that indicated by earlier observations by \citet{bash08}.
\subsection{Mass limits from accretion models}
\bildan{}{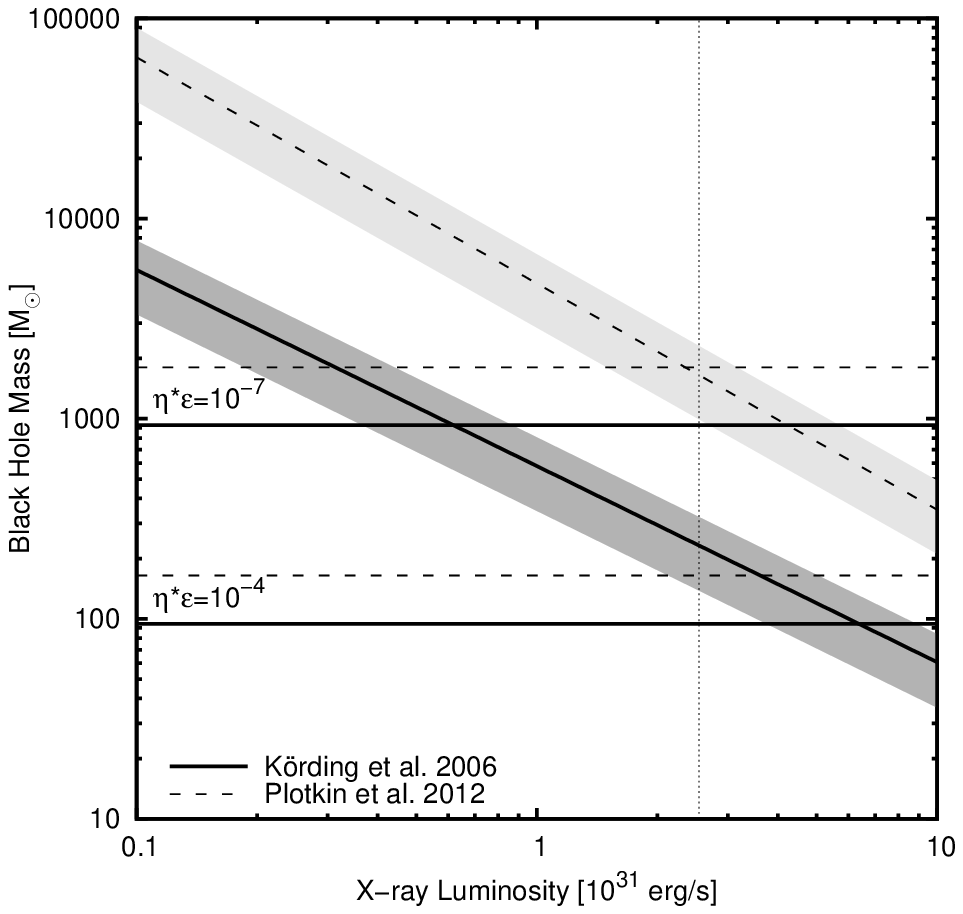}
 {Black hole mass as a function of X-ray luminosity for our measured radio luminosity ($L_R=6.34\times10^{27}$ erg/s) as predicted by the FP from \citet{kording06} (solid lines) and
   \citet{plotkin12} (dashed lines). The gray shaded areas indicate the 1$\sigma$ uncertainties of the relations. The horizontal lines mark
   the predicted masses for different assumptions on the radiative efficiency $\eta$ and the accretion efficiency $\epsilon$. The vertical dotted line marks the upper limit for
   the X-ray luminosity from \citet{hannikainen05}.
 }
 {fig: fp_comp}
\bildaus

The lack of a central X-ray source in M15 is expected because the accretion process is most likely radiatively inefficient \citep{ho03}. To constrain the mass of the
IMBH--candidate we estimate the X-ray luminosity based on the (unknown) accretion rate $\dot M$:
\begin{equation*}\label{equ: xray mass accretion}
 L_X=\eta\epsilon \:c^{2}\dot M \;\;,
\end{equation*}
where $\eta$ and $\epsilon$ are the radiation efficiency and the accretion
efficiency, respectively. The accretion process of radiatively inefficient black hole binaries (e.g. \citeauthor{esin97} 1997) as well as those of quiescent SMBHs
(e.g. \citeauthor{yuan03} 2003) is well described by advection-dominated accretion flow
models (ADAF, \citeauthor{narayan94} 1994). Such BHs undergo quasi-spherical accretion and, thus, following \citet{maccarone04} we further assume that mass accretion of the
putative IMBH in M15 can be described by the Bondi-Hoyle-Lyttleton (BHL) formalism \citep{hoyle41,bondi44,ho03}
\begin{equation*}\label{equ: bhl accretion}
 \dot M_{\text{BHL}}=3.2\times 10^{17}\left(\frac{M_{\bullet}}{2000\:\text{M}_{\odot}}\right)^{2}\left(\frac{n}{0.2\:\text{cm}^{-3}}\right)\left(\frac{T}{10^4\text{K}}\right)^{-1.5}\text{g s}^{-1} \;.
\end{equation*}
Here, $n$ and $T$ are the gas density and temperature in the GC, respectively. We adopt $n=0.2$ \mbox{cm$^{-3}$} from \citet{freire01} and use a typical GC gas temperature of
$T=10^{4}$ K. Inserting the above equations into the FP, and solving for $M_{\bullet}$ yields
\begin{equation*}\label{equ: bh mass solved}
2.96\;\text{log}M_{\bullet}=1.55\;\text{log}L_R-0.98\;\text{log}(\eta*\epsilon) -41.17 \;.
\end{equation*}
Mass estimates from this relation depend crucially on the choice of $\eta$ and $\epsilon$. Observations of the IMBH--candiate in G1 in M31 suggest that the radiative efficiency
$\eta<0.01$ for this source \citep{ulvestad07}. This is consistent with the aforementioned ADAF models that are only valid for $\eta\ll0.1$ \citep{narayan08}. In order to cover as large a parameter
space as possible we choose $\eta=0.1$ as conservative upper limit for the radiative efficiency. As lower limit we take the estimate of \citet{ho03} who state
that $\eta$ can be as low as $10^{-4}$ for the IMBH candidate in M15. It is not clear at all at what fraction $\epsilon$ of the Bondi-rate the central source is accreting. Typical
values are in the range $\epsilon=[10^{-3}, 0.1]$ \citep{maccarone04, ulvestad07, cseh10} which we adopt for our analysis. Consequently, we explore a parameter space covering
$\eta*\epsilon = [0.01, 10^{-7}]$. Based on the upper X-ray luminosity as measured by \citet{hannikainen05} we can exclude values of $\eta*\epsilon\geq10^{-3}$ because
they would yield stronger X-ray emission than what is observed. The mass limit for a value of $\eta*\epsilon=10^{-4}$ is indicated in Fig. \ref{fig: fp_comp} and would yield a
black hole mass $M_{\bullet}=94\pm37$ \mbox{\msol}. For
$\eta*\epsilon=10^{-5}$ we find $M_{\bullet}=202\pm80$ \mbox{\msol} which would indicate a source that is accreting at ten percent of the Bondi-rate with a radiative efficiency\footnote{Obviously, other combinations of $(\eta,\epsilon)$ would also be suitable but we give this one as a limiting case that has been discussed by other authors
  (e.g. \citeauthor{ho03} 2003, \citeauthor{cseh10} 2010).}
 $\eta=10^{-4}$. The combination of both, very inefficient radiation ($\eta=10^{-4}$) and inefficient accretion ($\epsilon=10^{-3}$) is also indicated in Fig. \ref{fig: fp_comp} with
the mass estimate of $M_{\bullet}=927\pm371$ \mbox{\msol}. 

\subsection{Mass estimates from other FP--relations}
Mass estimates computed here with the FP--relation from \citet{kording06} agree well within the errors with those estimates computed from the relations found by, e.g., 
\citet{merloni03} and \citet{gultekin09}. We note, however, a significant offset towards higher masses when using the latest version of the FP published by \citet{plotkin12}. Their relation reads
\[
\text{log}M_{\bullet}=1.64\;\text{log}L_R-1.13\;\text{log}L_X-6.89 \;\;.
\]
In Fig. \ref{fig: fp_comp} the predicted BH mass from this relation for our measured radio flux density is indicated by the dashed lines. With this relation
the limit for the X-ray luminosity results in a BH mass $M_{\bullet}=1654\pm661$ \mbox{\msol}. However, this mass limit can only be explained by the least efficient accretion
limit of $\eta*\epsilon=10^{-7}$. All higher values of $\eta*\epsilon$ would yield an X-ray luminosity beyond the one observed and can be excluded within this FP--relation. 
\section{Conclusions}
Using our multi-epoch high sensitivity observations of M15 we were able to put an upper limit of 10 $\mu$Jy on the 1.6 GHz radio flux density of a central source in this globular
cluster. Assuming that the central mass concentration is a black hole we employed the FP as derived by \citet{kording06} to put constraints on the
mass of the source. The lack of a detection of an object in the X-ray observations coinciding with the assumed cluster center allowed us to use an upper limit for the X-ray luminosity of the putative IMBH which
yields a black hole mass $M_{\bullet}=232\pm93$ \mbox{\msol}. Such a mass estimate is in agreement with a source accreting matter from its surrounding at up to ten percent of the
Bondi-rate with a radiative efficiency as low as $10^{-4}$ which is consistent with results for the IMBH--candidate in G1 in M31 \citep{ulvestad07}. Even the $3\sigma$ upper limit of $\approx500$ \mbox{\msol}, however, is still a factor of seven lower than that
required to explain the dynamics of the cluster \citep{vandenbosch06}. If we employ the FP from \citet{plotkin12} the $3\sigma$ upper mass limit is increased to $\approx3600$
\mbox{\msol} which would agree with the required dynamical mass. However, such a mass can only be explained if the central object is accreting at 0.1 percent of the Bondi rate
and if
only 0.01 percent of radiation is escaping from the accretion region. Furthermore, this mass estimate disagrees by a factor of seven with results obtained from most other FP--relations and
we adopt the mass limit of $M_{\bullet}=232\pm93$ \mbox{\msol}.

Therefore, we conclude that M15 most likely does not
contain an IMBH but that its central region probably hosts a collection of dark
remants such as neutron stars as
proposed by, e.g., \citet{baumgardt03, mcnamara03}, and \citet{murphy11}
instead. The reason for not detecting these pulsars is most probably their
expected low flux density of $\approx2$ \mbox{$\mu$Jy} \citep{sun02}.

During the final stages of the refereeing process of this paper, \citet{strader12} published a mass estimate for the IMBH in M15 that is in good agreement with the estimate derived here.

\begin{acknowledgements}
The authors would like to thank the JIVE staff for correlating and for support
with scheduling the complex observations. We would also like to thank the anonymous referee for productive criticism. F.K. acknowledges partial support through the Bonn-Cologne Graduate School of Physics and Astronomy. This research is supported by the European Community Framework Programme 7, Advanced Radio Astronomy in Europe, grant agreement No. 227290.
\end{acknowledgements}
\bibliographystyle{aa}
\bibliography{Franz_Kirsten.bib}

\begin{thebibliography}{50}
\expandafter\ifx\csname natexlab\endcsname\relax\def\natexlab#1{#1}\fi

\bibitem[{{Bash} {et~al.}(2008){Bash}, {Gebhardt}, {Goss}, \& {Vanden
  Bout}}]{bash08}
{Bash}, F.~N., {Gebhardt}, K., {Goss}, W.~M., \& {Vanden Bout}, P.~A. 2008,
  \aj, 135, 182

\bibitem[{{Baumgardt} {et~al.}(2003){Baumgardt}, {Hut}, {Makino}, {McMillan},
  \& {Portegies Zwart}}]{baumgardt03}
{Baumgardt}, H., {Hut}, P., {Makino}, J., {McMillan}, S., \& {Portegies Zwart},
  S. 2003, \apjl, 582, L21

\bibitem[{{Berghea} {et~al.}(2008){Berghea}, {Weaver}, {Colbert}, \&
  {Roberts}}]{berghea08}
{Berghea}, C.~T., {Weaver}, K.~A., {Colbert}, E.~J.~M., \& {Roberts}, T.~P.
  2008, \apj, 687, 471

\bibitem[{{Bondi} \& {Hoyle}(1944)}]{bondi44}
{Bondi}, H. \& {Hoyle}, F. 1944, \mnras, 104, 273

\bibitem[{{Clark} {et~al.}(1975){Clark}, {Markert}, \& {Li}}]{clark75}
{Clark}, G.~W., {Markert}, T.~H., \& {Li}, F.~K. 1975, \apjl, 199, L93

\bibitem[{{Colbert} \& {Mushotzky}(1999)}]{colbert99}
{Colbert}, E.~J.~M. \& {Mushotzky}, R.~F. 1999, \apj, 519, 89

\bibitem[{{Cseh} {et~al.}(2010){Cseh}, {Kaaret}, {Corbel}, {K{\"o}rding},
  {Coriat}, {Tzioumis}, \& {Lanzoni}}]{cseh10}
{Cseh}, D., {Kaaret}, P., {Corbel}, S., {et~al.} 2010, \mnras, 406, 1049

\bibitem[{{Djorgovski} \& {King}(1986)}]{djorgovski86}
{Djorgovski}, S. \& {King}, I.~R. 1986, \apjl, 305, L61

\bibitem[{{Esin} {et~al.}(1997){Esin}, {McClintock}, \& {Narayan}}]{esin97}
{Esin}, A.~A., {McClintock}, J.~E., \& {Narayan}, R. 1997, \apj, 489, 865

\bibitem[{{Falcke} {et~al.}(2004){Falcke}, {K{\"o}rding}, \&
  {Markoff}}]{falcke04}
{Falcke}, H., {K{\"o}rding}, E., \& {Markoff}, S. 2004, \aap, 414, 895

\bibitem[{{Farrell} {et~al.}(2009){Farrell}, {Webb}, {Barret}, {Godet}, \&
  {Rodrigues}}]{farrell09}
{Farrell}, S.~A., {Webb}, N.~A., {Barret}, D., {Godet}, O., \& {Rodrigues},
  J.~M. 2009, \nat, 460, 73

\bibitem[{{Ferrarese} \& {Merritt}(2000)}]{ferrarese00}
{Ferrarese}, L. \& {Merritt}, D. 2000, \apjl, 539, L9

\bibitem[{{Freire} {et~al.}(2001){Freire}, {Kramer}, {Lyne}, {Camilo},
  {Manchester}, \& {D'Amico}}]{freire01}
{Freire}, P.~C., {Kramer}, M., {Lyne}, A.~G., {et~al.} 2001, \apjl, 557, L105

\bibitem[{{Gebhardt} {et~al.}(2000){Gebhardt}, {Bender}, {Bower}, {Dressler},
  {Faber}, {Filippenko}, {Green}, {Grillmair}, {Ho}, {Kormendy}, {Lauer},
  {Magorrian}, {Pinkney}, {Richstone}, \& {Tremaine}}]{gebhardt00}
{Gebhardt}, K., {Bender}, R., {Bower}, G., {et~al.} 2000, \apjl, 539, L13

\bibitem[{{Gebhardt} {et~al.}(2002){Gebhardt}, {Rich}, \& {Ho}}]{gebhardt02}
{Gebhardt}, K., {Rich}, R.~M., \& {Ho}, L.~C. 2002, \apjl, 578, L41

\bibitem[{{Gerssen} {et~al.}(2003){Gerssen}, {van der Marel}, {Gebhardt},
  {Guhathakurta}, {Peterson}, \& {Pryor}}]{gerssen03}
{Gerssen}, J., {van der Marel}, R.~P., {Gebhardt}, K., {et~al.} 2003, \aj, 125,
  376

\bibitem[{{Giacconi} {et~al.}(1974){Giacconi}, {Murray}, {Gursky}, {Kellogg},
  {Schreier}, {Matilsky}, {Koch}, \& {Tananbaum}}]{giacconi74}
{Giacconi}, R., {Murray}, S., {Gursky}, H., {et~al.} 1974, \apjs, 27, 37

\bibitem[{{Goldsbury} {et~al.}(2010){Goldsbury}, {Richer}, {Anderson},
  {Dotter}, {Sarajedini}, \& {Woodley}}]{goldsbury10}
{Goldsbury}, R., {Richer}, H.~B., {Anderson}, J., {et~al.} 2010, \aj, 140, 1830

\bibitem[{{G{\"u}ltekin} {et~al.}(2009){G{\"u}ltekin}, {Cackett}, {Miller}, {Di
  Matteo}, {Markoff}, \& {Richstone}}]{gultekin09}
{G{\"u}ltekin}, K., {Cackett}, E.~M., {Miller}, J.~M., {et~al.} 2009, \apj,
  706, 404

\bibitem[{{Hannikainen} {et~al.}(2005){Hannikainen}, {Charles}, {van Zyl},
  {Kong}, {Homer}, {Hakala}, {Naylor}, \& {Davies}}]{hannikainen05}
{Hannikainen}, D.~C., {Charles}, P.~A., {van Zyl}, L., {et~al.} 2005, \mnras,
  357, 325

\bibitem[{{Ho} {et~al.}(2003){Ho}, {Terashima}, \& {Okajima}}]{ho03}
{Ho}, L.~C., {Terashima}, Y., \& {Okajima}, T. 2003, \apjl, 587, L35

\bibitem[{{Hoyle} \& {Lyttleton}(1941)}]{hoyle41}
{Hoyle}, F. \& {Lyttleton}, R.~A. 1941, \mnras, 101, 227

\bibitem[{{Illingworth} \& {King}(1977)}]{illingworth77}
{Illingworth}, G. \& {King}, I.~R. 1977, \apjl, 218, L109

\bibitem[{{Jacoby} {et~al.}(2006){Jacoby}, {Cameron}, {Jenet}, {Anderson},
  {Murty}, \& {Kulkarni}}]{jacoby06}
{Jacoby}, B.~A., {Cameron}, P.~B., {Jenet}, F.~A., {et~al.} 2006, \apjl, 644,
  L113

\bibitem[{{Johnston} {et~al.}(1991){Johnston}, {Kulkarni}, \&
  {Goss}}]{johnston91}
{Johnston}, H.~M., {Kulkarni}, S.~R., \& {Goss}, W.~M. 1991, \apjl, 382, L89

\bibitem[{{K{\"o}rding} {et~al.}(2006){K{\"o}rding}, {Falcke}, \&
  {Corbel}}]{kording06}
{K{\"o}rding}, E., {Falcke}, H., \& {Corbel}, S. 2006, \aap, 456, 439

\bibitem[{{Kormendy} \& {Richstone}(1995)}]{kormendy95}
{Kormendy}, J. \& {Richstone}, D. 1995, \araa, 33, 581

\bibitem[{{Lu} \& {Kong}(2011)}]{lu11}
{Lu}, T.-N. \& {Kong}, A.~K.~H. 2011, \apjl, 729, L25

\bibitem[{{Maccarone}(2004)}]{maccarone04}
{Maccarone}, T.~J. 2004, \mnras, 351, 1049

\bibitem[{{McClintock} \& {Remillard}(2006)}]{mcclintock06}
{McClintock}, J.~E. \& {Remillard}, R.~A. 2006, {Black hole binaries, Compact
  stellar X-ray sources}, ed. {Lewin, W.~H.~G.~\& van der Klis, M. (Cambridge
  University Press)}, 157--213

\bibitem[{{McNamara} {et~al.}(2003){McNamara}, {Harrison}, \&
  {Anderson}}]{mcnamara03}
{McNamara}, B.~J., {Harrison}, T.~E., \& {Anderson}, J. 2003, \apj, 595, 187

\bibitem[{{Merloni} {et~al.}(2003){Merloni}, {Heinz}, \& {di
  Matteo}}]{merloni03}
{Merloni}, A., {Heinz}, S., \& {di Matteo}, T. 2003, \mnras, 345, 1057

\bibitem[{{Miller} \& {Hamilton}(2002)}]{miller02}
{Miller}, M.~C. \& {Hamilton}, D.~P. 2002, \mnras, 330, 232

\bibitem[{{Murphy} {et~al.}(2011){Murphy}, {Cohn}, \& {Lugger}}]{murphy11}
{Murphy}, B.~W., {Cohn}, H.~N., \& {Lugger}, P.~M. 2011, \apj, 732, 67

\bibitem[{{Narayan} \& {McClintock}(2008)}]{narayan08}
{Narayan}, R. \& {McClintock}, J.~E. 2008, \nar, 51, 733

\bibitem[{{Narayan} \& {Yi}(1994)}]{narayan94}
{Narayan}, R. \& {Yi}, I. 1994, \apjl, 428, L13

\bibitem[{{Newell} {et~al.}(1976){Newell}, {Da Costa}, \& {Norris}}]{newell76}
{Newell}, B., {Da Costa}, G.~S., \& {Norris}, J. 1976, \apjl, 208, L55

\bibitem[{{{\"O}zel} {et~al.}(2010){{\"O}zel}, {Psaltis}, {Narayan}, \&
  {McClintock}}]{ozel10}
{{\"O}zel}, F., {Psaltis}, D., {Narayan}, R., \& {McClintock}, J.~E. 2010,
  \apj, 725, 1918

\bibitem[{{Plotkin} {et~al.}(2012){Plotkin}, {Markoff}, {Kelly}, {K{\"o}rding},
  \& {Anderson}}]{plotkin12}
{Plotkin}, R.~M., {Markoff}, S., {Kelly}, B.~C., {K{\"o}rding}, E., \&
  {Anderson}, S.~F. 2012, \mnras, 419, 267

\bibitem[{{Schilizzi} {et~al.}(2001){Schilizzi}, {Aldrich}, {Anderson}, {Bos},
  {Campbell}, {Canaris}, {Cappallo}, {Casse}, {Cattani}, {Goodman}, {van
  Langevelde}, {Maccafferri}, {Millenaar}, {Noble}, {Olnon}, {Parsley},
  {Phillips}, {Pogrebenko}, {Smythe}, {Szomoru}, {Verkouter}, \&
  {Whitney}}]{schilizzi01}
{Schilizzi}, R.~T., {Aldrich}, W., {Anderson}, B., {et~al.} 2001, Experimental
  Astronomy, 12, 49

\bibitem[{{Strader} {et~al.}(2012){Strader}, {Chomiuk}, {Maccarone},
  {Miller-Jones}, {Seth}, {Heinke}, \& {Sivakoff}}]{strader12}
{Strader}, J., {Chomiuk}, L., {Maccarone}, T., {et~al.} 2012, ArXiv e-prints
  (arXiv:1203.6352)

\bibitem[{{Sun} {et~al.}(2002){Sun}, {Han}, \& {Qiao}}]{sun02}
{Sun}, X.-H., {Han}, J.-L., \& {Qiao}, G.-J. 2002, \cjaa, 2, 133

\bibitem[{{Ulvestad} {et~al.}(2007){Ulvestad}, {Greene}, \& {Ho}}]{ulvestad07}
{Ulvestad}, J.~S., {Greene}, J.~E., \& {Ho}, L.~C. 2007, \apjl, 661, L151

\bibitem[{{van den Bosch} {et~al.}(2006){van den Bosch}, {de Zeeuw},
  {Gebhardt}, {Noyola}, \& {van de Ven}}]{vandenbosch06}
{van den Bosch}, R., {de Zeeuw}, T., {Gebhardt}, K., {Noyola}, E., \& {van de
  Ven}, G. 2006, \apj, 641, 852

\bibitem[{{van der Marel} \& {Anderson}(2010)}]{vandermarel10}
{van der Marel}, R.~P. \& {Anderson}, J. 2010, \apj, 710, 1063

\bibitem[{{Vesperini} {et~al.}(2010){Vesperini}, {McMillan}, {D'Ercole}, \&
  {D'Antona}}]{vesperini10}
{Vesperini}, E., {McMillan}, S.~L.~W., {D'Ercole}, A., \& {D'Antona}, F. 2010,
  \apjl, 713, L41

\bibitem[{{Walker} \& {Chatterjee}(1999)}]{walker99}
{Walker}, R.~C. \& {Chatterjee}, S. 1999, VLBA Scientific Memo 23 (Socorro, NM:
  NRAO), \url{http://www.vlba.nrao.edu/memos/sci/}

\bibitem[{{White} \& {Angelini}(2001)}]{white01}
{White}, N.~E. \& {Angelini}, L. 2001, \apjl, 561, L101

\bibitem[{{Yuan} {et~al.}(2003){Yuan}, {Quataert}, \& {Narayan}}]{yuan03}
{Yuan}, F., {Quataert}, E., \& {Narayan}, R. 2003, \apj, 598, 301

\bibitem[{{Zampieri} \& {Roberts}(2009)}]{zampieri09}
{Zampieri}, L. \& {Roberts}, T.~P. 2009, \mnras, 400, 677

\end{thebibliography}
\end{document}